# On Measurement Bias in Causal Inference

Judea Pearl
Cognitive Systems Laboratory
Computer Science Department
University of California, Los Angeles, CA 90024 USA
*judea@cs.ucla.edu*

## Abstract

This paper addresses the problem of measurement errors in causal inference and highlights several algebraic and graphical methods for eliminating systematic bias induced by such errors. In particulars, the paper discusses the control of partially observable confounders in parametric and non parametric models and the computational problem of obtaining bias-free effect estimates in such models.

## 1 INTRODUCTION

This paper discusses methods of dealing with measurement errors in the context of graph-based causal inference. It is motivated by a result known to epidemiologists (Greenland and Kleinbaum, 1983; Greenland and Lash, 2008) and regression analysts (Carroll et al., 2006; Selén, 1986) that has not been fully utilized in causal analysis or graphical models.

Consider the problem of estimating the causal effect of $X$ on $Y$ when a sufficient set $Z$ of confounders can only be measured with error (see Fig. 1), via a proxy set $W$. Since $Z$ is assumed sufficient, the causal effect is identified from measurement on $X, Y$, and $Z$, and can be written

$$P(y|do(x)) = \sum_z P(y|x,z)P(z)$$
$$= \sum_z P(x,y,z)/P(x|z) \qquad (1)$$

However, if $Z$ is unobserved, and $W$ is but a noisy measurement of $Z$, d-separation tells us immediately that adjusting for $W$ is inadequate, for it leaves the back-door path(s) $X \leftarrow Z \rightarrow Y$ unblocked.[1] Therefore, regardless of sample size, the effect of $X$ on $Y$ cannot be estimated without bias. It turns out, however, that if we are given the conditional probabilities $P(w|z)$ that govern the error mechanism we can perform a modified-adjustment for $W$ that, in the limit of a very large sample, would amount to the same thing as observing and adjusting for $Z$ itself, thus rendering the causal effect identifiable.

[1] For concise definitions and descriptions of graphical concepts such as "d-separation," "back-door," and "sufficiency" see (Pearl, 2009, pp. 335–6, 344–5).

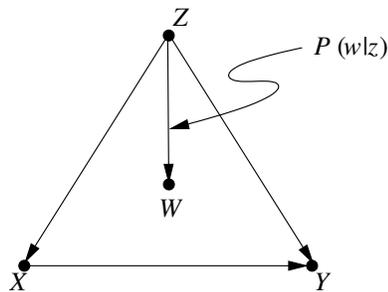

Figure 1: Needed the causal effect of $X$ on $Y$ when $Z$ is unobserved, and $W$ provides a noisy measurement of $Z$.

The possibility of removing bias by modified adjustment is far from obvious, because, although $P(w|z)$ is assumed given, the actual value of the confounder $Z$ remains uncertain for each measurement $W = w$, so one would expect to get either a distribution over causal effects, or bounds thereof. Instead, we can actually get a repaired point estimate of $P(y|do(x))$ that is asymptotically unbiased.

This result, which I will label "effect restoration," has powerful consequences in practice because, when $P(w|z)$ is not given, one can resort to a Bayesian (or bounding) analysis and assume a prior distribution (or bounds) on the parameters of $P(w|z)$ which would yield a distribution (or bounds) over $P(y|do(x))$ (Greenland, 2007). Alternatively, if costs permit, one can estimate $P(w|z)$ by re-testing $Z$ in a sampled sub-population (Carroll et al., 2006).[2]

[2] In the literature on measurement errors and sensitiv-

On the surface, the possibility of correcting for measurement bias seems to undermine the importance of accurate measurements. It suggests that as long as we know how bad our measurements are there is no need to correct them because they can be corrected post-hoc by analytical means. This is not so. First, although an unbiased effect estimate can be recovered from noisy measurements, sampling variability increases substantially with error. Second, even assuming unbounded sample size, the estimate will be biased if the postulated $P(w|z)$ is incorrect.[3]

Effect restoration can be analyzed from either a statistical or causal viewpoint. Taking the statistical view, one may argue that, once the effect $P(y|do(x))$ is identified in terms of a latent variable $Z$ and given by the estimand in (1), the problem is no longer one of causal inference, but rather of regression analysis, whereby the regressional expression $E_z P(y|x, z)$ need to be estimated from a noisy measurement of $Z$, as given by $W$. This is indeed the approach taken in the vast literature on measurement error (e.g., (Selén, 1986; Carroll et al., 2006)).

The causal analytic perspective is different; it maintains that the ultimate purpose of the analysis is not the statistics of $X, Y$, and $Z$, as is normally assumed in the measurement literature, but a causal quantity $P(y|do(x))$ that is mapped into regression vocabulary only when certain causal assumptions are deemed plausible. Moreover, the very idea of modeling the error mechanism $P(w|z)$ requires causal considerations; errors caused by noisy measurements are fundamentally different from those caused by noisy agitators. Indeed, the reason we seek an estimate $P(w|z)$ as opposed to $P(z|w)$, be it from judgment or from pilot studies, is that we consider the former parameter to be a more reliable and transportable than the latter. Transportability is a causal consideration that is hardly touched upon in the statistical measurement literature, where causal vocabulary is usually avoided and causal relations relegated to informal judgment (e.g., (Carroll et al., 2006, pp. 29–32)).

Viewed from this perspective, the measurement error literature appears to be engaged (unwittingly) in a causal inference exercise that can benefit substantially from making the causal framework explicit. The benefit can in fact be mutual; identifiability with partially specified causal parameters (as in Fig. 1) is rarely discussed in the causal inference literature (notable exceptions are Hernán and Cole (2009), Goetghebeur and Vansteelandt (2005), and Cai and Kuroki (2008)), while graphical models are hardly used in the measurement error literature.

In this paper we will consider the mathematical aspects of effect restoration and will focus on asymptotic analysis. Our aims are to understand the conditions under which effect restoration is feasible, to assess the computational problems it presents, and to identify those features of $P(w|z)$ and $P(x, y, w)$ that are major contributors to measurement bias, and those that contribute to robustness against bias.

## 2 EFFECT RESTORATION BY MATRIX ADJUSTMENT

The main idea, adapted from Greenland and Lash (2008, p. 360), is as follows: Starting with the joint probability $P(x, y, z, w)$, and assuming that $W$ depends only on $Z$,[4] i.e.,

$$P(w|x, y, z) = P(w|z) \qquad (2)$$

we write

$$\begin{aligned} P(x, y, w) &= \sum_z P(x, y, z, w) \\ &= \sum_z P(w|x, y, z) P(x, y, z) \\ &= \sum_z P(w|z) P(x, y, z) \end{aligned}$$

For each $x$ and $y$, we can interpret the transformation above as a vector-matrix multiplication:

$$V(w) = \sum_z M(w, z) V(z)$$

where $V(w) = P(x, y, w)$ and $M(w, z) = P(w|z)$ is a stochastic matrix (i.e., the entries in each column are non-negative and sum to one). It is well known that, under fairly broad conditions, $M$ has an inverse (call it $I$), which allows us to write:

$$P(x, y, z) = \sum_w I(z, w) P(x, y, w) \qquad (3)$$

We are done now, because (3) enables us to reconstruct the joint distribution of $X, Y$, and $Z$ from that of the

---

ity analysis, this sort of exercise is normally done by recalibration techniques (Greenland and Lash, 2008). The latter employs a "validation study" in which $Z$ is measured without error in a subpopulation and used to calibrate the estimates in the main study (Selén, 1986).

[3]In extreme cases, wrongly postulated $P(w|z)$ may conflict with the data, and no estimate will be obtained. For example, if we postulate a non informative $W$, $P(w|z) = P(w)$, and we find that $W$ strongly depends on $X$, a contradiction arises and no effect estimate will emerge.

[4]This assumption goes under the rubric: "non-differential error" (Carroll et al., 2006).

observed variables, $X, Y$, and $W$. Thus, each term on the right hand side of (1) can be obtained from $P(x, y, w)$ through (3) and, assuming $Z$ is a sufficient set (i.e., satisfying the back-door test), $P(y|do(x))$ is estimable from the available data. Explicitly, we have:

$$\begin{aligned} P(y|do(x)) &= \sum_z P(y,z,x)P(z)/P(x,z) \\ &= \sum_z \sum_w I(z,w)P(x,y,w) \\ &\quad \frac{\sum_{xyw} I(z,w)P(x,y,w)}{\sum_{wy} I(z,w)P(x,y,w)} \\ &= \sum_z \sum_w I(z,w)P(x,y,w) \\ &\quad \frac{\sum_w I(z,w)P(w)}{\sum_w I(z,w)P(x,w)} \end{aligned} \quad (4)$$

Note that the same inverse matrix, $I$, appears in all summations. This will not be the case when we do not assume independent noise mechanisms. In other words, if (2) does not hold, we must write:

$$\begin{aligned} P(x,y,w) &= \sum_z P(w|x,y,z)P(x,y,z) \\ &= \sum_z M_{xy}(w,z)P(x,y,z) \end{aligned}$$

where $M_{xy}$ and its inverse $I_{xy}$ are both indexed by the specific values of $x$ and $y$, and we then obtain:

$$P(x,y,z) = \sum_w I_{xy}(z,w)P(x,y,w) \quad (5)$$

which, again, permits the identification of the causal effect via (4) except that the expression becomes somewhat more complicated. It is also clear that errors in the measurement of $X$ and $Y$ can be absorbed into a vector $W$, and do not present any conceptual problem.

Equation (4) demonstrates the feasibility of effect reconstruction and proves that, despite the uncertainty in the variables $X, Y$ and $Z$, the causal effect is identifiable once we know the statistics of the error mechanism.[5]

This result is reassuring, but presents practical challenges of both representation, computation and estimation. Given the potentially high dimensionality of $Z$ and $W$, the parameterization of $I$ would in general be impractical or prohibitive. However, if we can assume independent local mechanisms, $P(w|z)$ can be decomposed into a product $P(w|z) =$ $P(w_1|z_1)P(w_2|z_2), \ldots, P(w_k|z_k)$ which renders $I$ decomposable as well. Even when full decomposition is not plausible, sparse couplings between the different noise mechanisms would enable parsimonious parameterization using, for example, Bayesian networks.

The second challenge concerns the summations in Eq. (4) which, taken literally, calls for exponentially long summation over all values of $w$. In practice, however, this can be mitigated since, for any given $z$, there will be only small number of w's for which $I(z,w)$ is non-negligible. This computation, again, can be performed efficiently using Bayes networks inference.

This still would not permit us to deal with the problem of empty cells which, owed to the high dimensionality of $Z$ and $W$ would prevent us from getting reliable statistics of $P(x,y,w)$, as required by (4). One should resort therefore to propensity score ($PS$) methods, which map the cells of $Z$ onto a single scalar.

The error-free propensity score $L(z) = P(X=1|Z=z)$ being a functional of $P(x,y,z)$ can of course be estimated consistently from samples of $P(x,y,w)$ using the transformation (3). Explicitly, we have:

$$\begin{aligned} L(z) &= P(X=1|Z=z) \\ &= P(X=1, Z=z)/P(z) \\ &= \sum_y P(X=1, y, z)/\sum_{xy} P(x,y,z) \end{aligned}$$

where $P(x,y,z)$ is given in (3).

Using the decomposition in (2), we can further write:

$$\begin{aligned} L(z) &= \sum_y P(X=1, y, z)/\sum_{xy} P(x,y,z) \\ &= \sum_w I(z,w)P(X=1,w)/\sum_w I(z,w)P(w) \quad (6) \\ &= \sum_w I(z,w)L(w)P(w)/\sum_w I(z,w)P(w) \end{aligned}$$

where $L(w)$ is the error-prone propensity score

$$L(w) = P(X=1|W=w).$$

We see that $L(z)$ can be computed from $I(z,w)$, $L(w)$ and $P(w)$. Thus, if we succeed in estimating these three quantities in a parsimonious parametric form, the computation of $L(z)$ would be hindered only by the summations called for in (5). Once we estimate $L(w)$ parametrically for each conceivable $w$, Eq. (6) permits us to assign to each tuple $z$ a bias-less score $L(z)$ that correctly represents the probability of $X = 1$ given $Z = z$. This, in turn, should permit us to estimate, for each stratum $L = l$, the probability

$$P(l) = \sum_{z|L(z)=l} P(z)$$

---

[5]Kuroki and Pearl (2010) have further shown that, under certain conditions, the matrix $P(w|z)$ can be inferred from the data alone.

then compute the causal effect using

$$P(y|do(x)) = \sum_l P(y|x,l)P(l).$$

One technique for approximating $P(l)$ was proposed by Stürmer et al. (2005), and Schneeweiss et al. (2009) which does not make full use of the inversion in (9) or of graphical methods facilitating this inversion. A more promising approach would be to construct $P(l)$ and $P(y|x,l)$ directly from synthetic samples of $P(x,y,z)$ that can be created to mirror the empirical samples of $P(x,y,w)$. This is illustrated in the next subsection, using binary variables.

## 3 EFFECT RESTORATION IN BINARY MODELS

Let $X, Y, Z$ and $W$ be binary variables, and let the noise mechanism be characterizes by

$$P(W=0|Z=1) = \epsilon$$
$$P(W=1|Z=0) = \delta$$

To simplify notation, let the propositions $Z = 1$ and $Z = 0$ be denoted by $z_1$ and $z_0$, respectively, and the same for $W = 1$ and $W = 0$, so that $\epsilon$ and $\delta$ can be written

$$\epsilon = P(w_0|z_1)$$
$$\delta = P(w_1|z_0)$$

Equation (3) then translates to

$$P(x,y,z_0) = \frac{(1-\epsilon)P(x,y,w_0) - \epsilon P(x,y,w_1)}{(1-\epsilon-\delta)}$$
$$P(x,y,z_1) = \frac{-\delta P(x,y,w_0) + (1-\delta)P(x,y,w_1)}{(1-\epsilon-\delta)} \quad (7)$$

which represents the inverse matrix

$$I(w,z) = \begin{bmatrix} 1-\delta & \epsilon \\ \delta & 1-\epsilon \end{bmatrix}^{-1}$$
$$= \frac{1}{1-\epsilon-\delta} \begin{bmatrix} 1-\epsilon & -\epsilon \\ -\delta & 1-\delta \end{bmatrix}$$

Metaphorically, the transformation in (7) can be described as a mass re-assignment process, as if every two cells, $(x, y, w_0)$ and $(x, y, w_1)$, compete on how to split their combined weight $P(x, y)$ between the two latent cells $(x, y, z_0)$ and $(x, y, z_1)$ thus creating a synthetic population $P(x, y, z)$ from which (4) follows. Figure 2 describes how $P(w_1|x, y)$, the fraction of the weight held by the $(x, y, w_1)$ cell determines the ratio $P(z_1|x, y)/P(z_0|x, y)$ of weights that is eventually received by cells $(x, y, z_1)$ and $(x, y, z_0)$, respectively. We see that when $P(w_1|x, y)$ approaches $1-\epsilon$, most of the $P(x, y)$ weight goes to cell $(z_1, x, y)$, whereas when $P(w_1|x, y)$ approaches $\delta$, most of that weight the twin cell $(z_0, x, y)$.

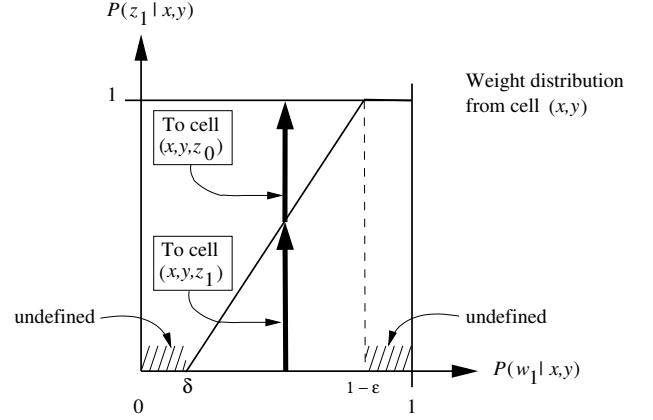

Figure 2: A curve describing how the weight $P(x, y)$ is distributed to cells $(x, y, z_1)$ and $(x, y, z_0)$, as a function of $P(w_1|x, y)$.

Clearly, when $\epsilon + \delta = 1$, $W$ provides no information about $Z$ and the inverse does not exist. Likewise, whenever any of the synthetic probabilities $P(x, y, z)$ falls outside the $(0, 1)$ interval, a modeling constraint is violated (see (Pearl, 1988, Chapter 8)) meaning that the observed distribution $P(x, y, w)$ and the postulated error mechanism $P(w|z)$ are incompatible with the structure of Fig. 1 (see footnote 3). If we assign reasonable priors to $\epsilon$ and $\delta$, the linear function in Fig. 2 will become an $S$-shaped curve over the entire $[0, 1]$ interval, and each sample $(x, y, w)$ can then be used to update the relative weight $P(x, y, z_1)/P(x, y, z_0)$.

To compute the causal effect $P(y|do(x))$ we need only substitute $P(x, y, z)$ from (7) into Eqs. (1) or (4), which gives (after some algebra)

$$P(y|do(x)) = \frac{P(x,y,w_1)}{P(x|w_1)} \frac{\left[1 - \frac{\delta}{P(w_1|x,y)}\right]\left[1 - \frac{\delta}{P(w_1)}\right]}{1 - \delta P(x)/P(w_1)}$$
$$+ \frac{P(x,y,w_0)}{P(x|w_0)} \frac{\left[1 - \frac{\epsilon}{P(w_0|x,y)}\right]\left[1 - \frac{\epsilon}{P(w_0)}\right]}{1 - \epsilon P(x)/P(w_0)}.$$
(8)

This expression highlights the difference between the standard and modified adjustment for $W$; the former (Eq. (1)), which is valid if $W = Z$, is given by the standard inverse probability weighting (e.g., (Pearl, 2009, Eq. (3.11))):

$$P(y|do(x)) = \frac{P(x,y,w_1)}{P(x|w_1)} + \frac{P(x,y,w_0)}{P(x|w_0)}$$

The extra factors in Eq. (8) can be viewed as modifiers of the inverse probability weight needed for a bias-free estimate. Alternatively, these terms can be used to assess, given $\epsilon$ and $\delta$, what bias would be introduced if we ignore errors altogether and treat $W$ as a faithful representation of $Z$.

The infinitesimal approximation of (8), in the limit $\epsilon \to 0, \delta \to 0$, reads:

$$P(y|do(x))$$
$$\cong \frac{P(x,y,w_1)}{P(x|w_1)} \left[1 - \delta\left(\frac{1}{P(w_1|x,y)} - \frac{1-P(x)}{P(w_1)}\right)\right]$$
$$+ \frac{P(x,y,w_0)}{P(x|w_0)} \left[1 - \epsilon\left(\frac{1}{P(w_0|x,y)} - \frac{1-P(x)}{P(w_0)}\right)\right]$$

We see that, even with two error parameters ($\epsilon$ and $\delta$), and eight cells, the expression for $P(y|do(x))$ does not simplify to provide an intuitive understanding of the effect of $\epsilon$ and $\delta$ on the estimand. Such evaluation will be facilitated in linear models (Section 4).

Assuming now that $Z$ is a sufficient set of $K$ binary variables and, similarly, $W$ is a set of $K$ local indicators of $Z$, satisfying (2). Each sample $(x, y, w)$ should give rise to a synthetic distribution over the $2^K$ cells of $(x, y, z)$ given by a product of $K$ local distributions in the form of (7). This synthetic distribution can be sampled to generate synthetic $(x, y, z)$ samples, from which the true propensity score $L(z) = P(X=1|z)$ as well as the causal effect $P(y|do(x))$ can be estimated, as discussed in Section 2.

## 4 EFFECT RESTORATION IN LINEAR MODELS

Figure 3 depicts a linear version of the structural equation model (SEM) shown in Fig. 1. Here, the task is to estimate the effect coefficient $c_0$, while the parameters $c_3$ and $var(\epsilon_w)$, representing the noise mechanism $W = c_3 Z + \epsilon_W$, are assumed given.

Linear models offer two advantageous in handling measurement errors. First, they provide a more transparent picture into the role of each factor in the model. Second, certain aspects of the error mechanism can often be identified without resorting to external studies. This occurs, for example, when $Z$ possesses two independent indicators, say $W$ and $V$ (as in Fig. 3(b)), in which case the product $c_3^2 var(Z)$ is identifiable and is given by:

$$c_3^2 var(Z) = \frac{cov(XW)cov(XV)}{cov(WV)}. \quad (9)$$

As we shall see below, this product is sufficient for identifying $c_0$.

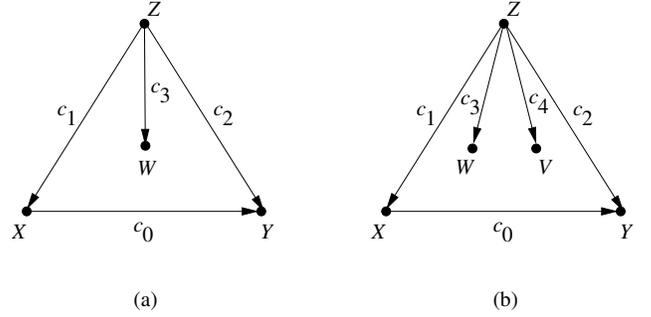

Figure 3: (a) A linear version of the model in Fig. 1. (b) A linear model with two indicators for $Z$, permitting the identification of $c_0$.

Equation (9) follows from Wright's rules of path analysis and reflects the well known fact (e.g., (Bollen, 1989, p. 224)) that, in linear models, structural parameters are identifiable (up to a constant $var(Z)$) whenever each latent variable (in our case $Z$) has three independent proxies (in our case $X, W$, and $V$)[6]

Cai and Kuroki (2008) and Kuroki and Pearl (2010) further showed that $c_0$ is identifiable from measurements of three proxies (of $Z$), even when these proxies are dependent of each other. For example, connecting $W$ to $X$ and $V$ to $Y$, still permits the identification of $c_0$.

To find $c_0$ in the model of Fig. 3, we write the three structural equations in the model

$$Y = c_2 Z + c_0 X + \epsilon_Y$$
$$W = c_3 Z + \epsilon_W$$
$$X = c_1 Z + \epsilon_X$$

and express the structural parameters in terms of the variances and covariances of the observed variables. This gives (after some algebra):

$$c_0 = \frac{cov(XY) - cov(XW)cov(WY)/c_3^2 var(Z)}{var(X) - cov^2(XW)/c_3^2 var(Z)} \quad (10)$$

and shows that the pivotal quantity needed for the identification of $c_0$ is the product

$$c_3^2 var(Z) = var(W) - var(\epsilon_W) \quad (11)$$

If we are in possession of several proxies for $Z$, $c_3^2 var(Z)$ can be estimated from the data, as in Eq. (9), yielding:

$$c_0 = \frac{cov(XY)cov(XV) - cov(YW)cov(WV)}{cov(XV)var(X) - cov(XW)cov(WV)} \quad (12)$$

---

[6]This partial identifiability of the so called "factor loadings," is not an impediment for the identification of $c_0$. However, if we were in possession of only one proxy (as in Fig. 3(a)) then knowledge of $c_3$ alone would be insufficient, the product $c_3^2 var(Z)$ is required.

If however $Z$ has only one proxy, $W$, as in Fig. 3(a), the product $c_3^2 var(Z)$ must be estimated externally, using either a pilot study or judgmental assessment.

The decomposition on the right hand side of Eq. (11) simplifies the judgmental assessment of $c_3^2 var(Z)$, since $c_3$ and $\epsilon_W$ are the parameters of an additive error mechanism
$$W = c_3 Z + \epsilon_W;$$
$c_3 = E(W|z)/z$ measures the slope with which the average of $W$ tracks the value of $Z$, while $var(\epsilon_W)$ measures the dispersion of $W$ around that average. $var(W)$ can, of course be estimated from the data.

Under a Gaussian distribution assumption, $c_3$ and $var(\epsilon_W)$ fully characterize the conditional density $f(w|z)$ which, according to Section 2, is sufficient for restoring the joint distribution of $x, y,$ and $z$, and thus secure the identification of the causal effect, through (1). This explains why the estimation of $c_3$ alone, be it from experimental data or our understanding of the physics behind the error process, is not sufficient for neutralizing the confounder $Z$. It also explains why the technique of "latent factor" analysis (Bollen, 1989) is sufficient for identifying causal effects, even though it fails to identify the "factor loading" $c_3$ separately of $var(Z)$.

In the noiseless case, i.e., $var(\epsilon_W) = 0$, we have $var(Z) = var(W)/c_3^2$ and Eq. (10) reduces to:
$$\begin{aligned} c_0 &= \frac{cov(XY) - cov(XW)cov(WY)/var(W)}{var(X) - cov^2(XW)/var(W)} \\ &= \frac{\beta_{yx} - \beta_{yw}\beta_{wx}}{1 - \beta_{xw}\beta_{wx}} \\ &= \beta_{yx\cdot w} \end{aligned} \quad (13)$$

where
$$\begin{aligned} \beta_{yx} &= cov(XY)/var(X) \\ &= \frac{\partial}{\partial x} E(Y|x) \end{aligned}$$

and $\beta_{yx\cdot w}$ is the coefficient of $x$ in the regression of $Y$ on $X$ and $W$, or:
$$\beta_{yx\cdot w} = \frac{\partial}{\partial x} E(Y|x, w)$$

As expected, the equality $c_0 = \beta_{yx\cdot z} = \beta_{yx\cdot w}$ assures a bias-free estimate of $c_0$ through adjustment for $W$, instead of $Z$; $c_3$ plays no role in this adjustment.

In the error-prone case, $c_0$ can be written
$$c_0 = \frac{\beta_{yx} - \beta_{yw}\beta_{wx}/k}{1 - \beta_{xw}\beta_{wx}/k}$$

where
$$k = 1 - var(\epsilon_W)/var(W)$$

and, as the formula reveals, $c_0$ cannot be interpreted in terms of an adjustment for a surrogate variable $V(W)$.

The strategy of adjusting for a surrogate variables has served as an organizing principle for many studies in traditional measurement error analysis (Carroll et al., 2006). For example, if one seeks to estimate the coefficient $c_1 = E(X|z)/z$ through a proxy $W$ of $Z$, one can always choose to regress $X$ on another variable, $V$, such that the slope of $X$ on $V$, $E(X|v)/v$, would yield an unbiased estimate of $c_1$. In our example of Fig. 3, one should choose $V$ to be the best linear estimate of $Z$, given $W$, namely $V = \alpha W$, where
$$\alpha = Cov(ZW)/var(W) = c_3 var(Z)/var(W)$$

is to be estimated separately, from a pilot study. However, this Two Stage Least Square strategy is not applicable in adjusting for latent confounders; i.e., there is no variable $V(W)$ such that $c_0 = \beta_{yx\cdot v}$.

## 5 MODEL TESTING WITH MEASUREMENT ERROR

When variables are measured without error, a structural equation model can be tested and diagnosed systematically by examining how well the data agrees with each statistical constraint that the model imposes on the joint distribution (or covariance matrix). The most common type of these constraints are conditional independence relations (or zero partial correlations), and these can be read off the causal diagram through the $d$-separation criterion (Pearl, 2009, pp. 335–7). For each missing edge in the diagram, say between $X$ and $Y$, the model dictates the conditional independence of $X$ and $Y$ given a set $Z$ of variables that $d$-separates $X$ from $Y$ in the diagram; these independencies can then be tested individually and systematically to assure compatibility between model and data before parameter identification commences (Kyono, 2010).

When $Z$ suffers from measurement errors (as in Fig. 1) those conditional independencies are not testable, since the proxies of $Z$ no longer $d$-separate $X$ from $Y$. The question arises whether surrogate tests exist through the available proxies, to detect possible violations of the missing-edge postulate. The preceding section suggests such tests, provided we know (or can estimate) the parameters of the error process.

This is seen by substituting $c_0 = 0$ in Eq. (10), and accepting the vanishing of the numerator as a surrogate test for $d$-separation between $X$ and $Y$:

**Theorem 1.** *If a latent variable $Z$ d-separates two measured variables, $X$ and $Y$, and $Z$ has a proxy $W$, $W = cZ + \epsilon_W$, then $cov(XY)$ must satisfy:*

$$cov(XY) = cov(XW)cov(WY)/c^2 var(Z)$$
$$= cov(XW)cov(WY)/[var(W) - var(\epsilon_W)] \quad (14)$$

We see that the usual condition of vanishing partial regression coefficient is replaced by a modified condition, in which $c^2 \, var(Z)$ needs to be estimated separately (as in Fig. 3(b)). If the product $c^2 var(Z)$ is estimated from other proxies of $Z$, as in Fig. 3(b), Eq. (14) assumes the form of a TETRAD condition (Bollen, 1989, p. 304).

$$cov(XY) = cov(VW)cov(WY)/cov(XV)$$

Cai and Kuroki (2008) derive additional conditions under which this constraint applies to multivariate sets of confounders and proxies.

Equation (14) can also be written

$$cov[Y(X - Wcov(XW)/\alpha)] = 0 \quad (15)$$

where $\alpha = var(W) - var(\epsilon_W)$, which provides an easy test of (14), in the style of Two Stage Least Square:

1. estimate $\alpha = var(W) - var(\epsilon_W)$ (using a pilot study or auxiliary proxy variables)

2. collect samples $X_i, Y_i, W_i \;\; i = 1, 2, 3, \ldots, n$

3. estimate $c_1 = cov(XW)$

4. Translate the data into fictitious samples $Y_i, V_i \;\; i = 1, 2, 3, \ldots, n$ with $V_i = X_i - c_1/\alpha W_i$

5. Compute (by Least Square) the best fit coefficient $a$ in $Y_i = aV_i + e_i$

6. Test if $a = 0$. If $a$ vanishes with sufficiently high confidence, then the data is compatible with the d-separation condition $X \perp\!\!\!\perp Y | Z$.

Theorem 1 can be generalized to include missing edges between latent variables, as well as between latent and observed variables. In fact, if the graph resulting from filling in a missing edge permits the identification of the corresponding edge coefficient $c$, then the original graph imposes a statistical constraint on the covariance matrix that can be used to test the absence of that edge. Such tests should serve as model-diagnostic tools, before (or instead) of submitting the entire model to a global test of fitness.

## 6 CONCLUSIONS

The paper discusses computational and representational problems connected with effect restoration when confounders are mismeasured or misclassified. In particular, we have explicated how measurement bias can be removed by creating synthetic samples from empirical samples, and how inverse-probability weighting can be modified to account for measurement error. Subsequently, we have analyzed measurement bias in linear systems and explicated graphical conditions under which such bias can be removed.

**Acknowledgments**

This note has benefited from discussions with Sander Greenland, Manabu Kuroki, Zhihong Cai, and Onye Arah and was supported in parts by grants from NIH #1R01 LM009961-01, NSF #IIS-0914211, and ONR #N000-14-09-1-0665.

## References

BOLLEN, K. (1989). *Structural Equations with Latent Variables*. John Wiley, New York.

CAI, Z. and KUROKI, M. (2008). On identifying total effects in the presence of latent variables and selection bias. In *Uncertainty in Artificial Intelligence, Proceedings of the Twenty-Fourth Conference* (D. McAllester and A. Nicholson, eds.). AUAI, Arlington, VA, 62–69.

CARROLL, R., RUPPERT, D., STEFANSKI, L. and CRAINICEANU, C. (2006). *Measurement Error in Nonlinear Models: A Modern Perspective*. 2nd ed. Chapman & Hall/CRC, Boca Raton, FL.

GOETGHEBEUR, E. and VANSTEELANDT, S. (2005). Structural mean models for compliance analysis in randomized clinical trials and the impact of errors on measures of exposure. *Statistical Methods in Medical Research* **14** 397–415.

GREENLAND, S. (2007). Bayesian perspectives for epidemiologic research. *International Journal of Epidemiology* **36** 195–202.

GREENLAND, S. and KLEINBAUM, D. (1983). Correcting for misclassification in two-way tables and matched-pair studies. *International Journal of Epidemiology* **12** 93–97.

GREENLAND, S. and LASH, T. (2008). Bias analysis. In *Modern Epidemiology* (K. Rothman, S. Greenland and T. Lash, eds.), 3rd ed. Lippincott Williams and Wilkins, Philadelphia, PA, 345–380.

Hernán, M. and Cole, S. (2009). Invited commentary: Causal diagrams and measurement bias. *American Journal of Epidemiology* **170** 959–962.

Kuroki, M. and Pearl, J. (2010). Measurement bias and effect restoration in causal inference. Tech. Rep. R-366, University of California, Los Angeles, CA. Forthcoming.

Kyono, T. (2010). Commentator: A front-end user-interface module for graphical and structural equation modeling. Tech. Rep. R-364, University of California, Los Angeles, CA. Master Thesis, <http://ftp.cs.ucla.edu/pub/stat_ser/r364.pdf>.

Pearl, J. (1988). *Probabilistic Reasoning in Intelligent Systems*. Morgan Kaufmann, San Mateo, CA.

Pearl, J. (2009). *Causality: Models, Reasoning, and Inference*. 2nd ed. Cambridge University Press, New York.

Schneeweiss, S., Rassen, J., Glynn, R., Avorn, J., Mogun, H. and Brookhart, M. (2009). High-dimensional propensity score adjustment in studies of treatment effects using health care claims data. *Epidemiology* **20** 512–522.

Selén, J. (1986). Adjusting for errors in classification and measurement in the analysis of partly and purely categorical data. *Journal of the American Statistical Association* **81** 75–81.

Stürmer, T., Schneeweiss, S., Avorn, J. and Glynn, R. (2005). Correcting effect estimates for unmeasured confounding in cohort studies with validation studies using propensity score calibration. *American Journal of Epidemiology* **162** 279–289.